\documentclass[12pt]{article}

\usepackage{graphicx}
\usepackage{cite}
\usepackage{amsmath,amssymb}
\usepackage{bm}
\usepackage{graphicx, color}
\usepackage{wrapfig}
\begin{document}
\title{
\begin{flushright}
\ \\*[-80pt] 
\begin{minipage}{0.2\linewidth}
\normalsize
%arXiv:YYMM.NNNN \\
KUNS-2380  \\*[50pt]
\end{minipage}
\end{flushright}
{\Large \bf 
Lepton flavor models with discrete prediction of 
$\theta_{13}$
\\*[20pt]}}

\author{
\centerline{
Hajime~Ishimori, \
Tatsuo~Kobayashi, \ }  \\
\\*[20pt]
\centerline{
\begin{minipage}{\linewidth}
\begin{center}
{\it \normalsize 
Department of Physics, Kyoto University, 
Kyoto 606-8502, Japan} \\
\end{center}
\end{minipage}}
\\*[50pt]}
\vskip 2 cm
\date{\small
\centerline{ \bf Abstract}
\begin{minipage}{0.9\linewidth}
\medskip 
 We study the lepton flavor models with the flavor 
symmetry $(Z_N \times Z_N \times Z_N)\rtimes Z_3$.
Our models predict non-vanishing discrete values of 
$\theta_{13}$ as well as $\theta_{12}$ and $\theta_{23}$ depending on $N$. 
For certain values, our models realize the tri-bimaximal mixing angles 
with $\theta_{13}=0$.
For other values, our models provide with discrete deviation from 
the  tri-bimaximal mixing angles.
\end{minipage}
}

\begin{titlepage}
\maketitle
\thispagestyle{empty}
\end{titlepage}

\section{Introduction}

It is the important issue to study the origin 
of the flavor structure in particle physics, 
that is, the hierarchy of quark/lepton masses and 
their mixing angles.
The form of lepton mixing angles is 
quite different from one of quark mixing angles 
(see e.g. for the lepton mixing 
\cite{Schwetz:2008er,Fogli:2008jx,GonzalezGarcia:2010er}).
The lepton mixing angles are large, while 
quark mixing angles are smaller.
The tri-bimaximal mixing Ansatz in the lepton sector 
is quite interesting \cite{Harrison:2002er} and the exact tri-bimaximal 
mixing leads to $\theta_{13}=0$.

Non-Abelian discrete flavor symmetry is a powerful tool to 
derive masses and mixing angles of quarks and leptons 
(see for review Refs. \cite{Altarelli:2010gt,Ishimori:2010au} 
and references therein). 
Especially, large mixing angles of leptons can naturally 
arise by assuming non-Abelian discrete flavor symmetries. 
Indeed, there are many models based on flavor symmetries to explain 
fermion masses and mixing.

Recently, the results of T2K and Double Chooz 
reported  a non-vanishing value of $\theta_{13}$ \cite{:2011sj,DeKerret-LowNu11}.
Thus, it is important to derive the lepton mixing 
with $\theta_{13} \neq 0$.
However, the tri-bimaximal mixing matrix is still 
good at a certain level.
Thus, it is interesting to study a small deviation, e.g. from 
the tri-bimaximal mixing angles with $\theta_{13} \neq 0$.
%Indeed, several works have been done.
%Here, we study the lepton flavor models, which lead to 
%deviations from the tri-bimaximal mixing angles by 
%discrete values.
%
%Non-Abelian discrete symmetry is a powerful tool to 
%predict masses and mixing angles of quark and lepton, 
%see review papers \cite{miller}-\cite{Ishimori:2010au}. 
%Especially, large mixing angles of lepton can naturally 
%arise by assuming non-Abelian discrete symmetries. 
%There are many models based on flavor symmetries to explain 
%fermion masses and mixing. 
After the result of T2K is released,
several attempts have been studied to 
to explain non-zero $U_{e3}$ with various methods 
\cite{Rodejohann:2011uz,Rashed:2011zs,Antusch:2011ic,Ding:2011gt,
Dev:2011hf,Araki:2011qy,Haba:2011nv,Blankenburg:2011vw,Marzocca:2011dh,
Deepthi:2011sk,Ludl:2011vv,Fritzsch:2011qv,He:2011hs,Ishimori:2011mt,
Xing:2011at,He:2011gb,Ma:2011yi,Zhou:2011nu,Araki:2011wn,Meloni:2011fx,
Morisi:2011pm,Chao:2011sp,Zhang:2011aw,Chu:2011jg,BhupalDev:2011gi,
Toorop:2011jn,Antusch:2011qg,deAdelhartToorop:2011re,Ding}.
For instance, introducing somewhat larger corrections to 
the model of tri-bimaximal mixing, one can obtain 
non-zero value for $U_{e3}$. However, such a method contains 
a free parameter so that it does not predict a specific value. 
In this paper, we build a model that predicts a special mixing 
pattern with  deviations from the tri-bimaximal mixing angles by 
discrete values, 
 which are consistent with experiments.

A simple way to derive the tri-bimaximal mixing is 
to assume residual symmetries for lepton mass matrices.  
Precisely, one can realize the tri-bimaximal mixing 
when there is the $Z_3$ symmetry for the charged lepton 
mass matrix
and the $Z_2$ symmetry for the neutrino mass matrix. 
These $Z_3$ and $Z_2$ symmetries may be originated from 
larger non-Abelian flavor symmetries like  
$A_4$, $S_4$, etc.
Thus, it is interesting to use non-Abelian discrete 
flavor symmetries 
such as $A_4$ and $S_4$ in order to derive the (exact) 
tri-bimaximal mixing \cite{Altarelli:2010gt,Ishimori:2010au}.
To predict non-vanishing $\theta_{13}$, we propose 
a modified symmetry for the charged leptons. 
In our approach, we use the
$(Z_N \times Z_N \times Z_N)\rtimes Z_3$ symmetry. 
With a choice of proper vacuum alignment, 
the charged lepton sector retains 
$Z_3 $ and $Z_N$ symmetries while the neutrino sector 
has $Z_2$ symmetry, which is the subgroup of $Z_N$. 
Due to the additional $Z_N$ symmetry of charged lepton, 
we obtain non-vanishing $\theta_{13}$.

%%%%%%%%%%

This paper is organized as follows. 
In section 2, we study about group-theoretical aspects of 
$(Z_N\times Z_N\times Z_N)\rtimes Z_3$
by giving conjugacy classes, character table, and tensor products.
In section 3 we introduce the model based on 
$(Z_N\times Z_N\times Z_N)\rtimes Z_3$ lepton flavor symmetry with 
one singlet and two triplet flavon fields. 
In section 4 we ensure the vacuum alignment assumed 
in the model by two ways, that is, 
analyzing scalar potential and using boundary conditions 
in extra dimensions.
In section 5, higher dimensional terms are taken into account.
Section 6 is devoted to the summary.

%%%%%%%%%%%%%%%%%%%%%%%%%%%%%%
%%%%%%%%%%%%%%%%%%%%%%%%%%%%%%

\section{$(Z_N \times Z_N \times Z_N)\rtimes Z_3$}

We use the group $(Z_N \times Z_N \times Z_N)\rtimes Z_3$ 
for the lepton flavor symmetry. 
Here, we study group-theoretical aspects of 
our group.
We denote the first, second and third $Z_N$ generators 
by $a$, $a'$ and $a"$, respectively, and the $Z_3$ generator 
is denoted by $b$.
They satisfy the following algebraic relations:
\begin{eqnarray}
\begin{split}
&a^N=a'^N=a''^N=b^3=e,
\quad
aa'=a'a,
\quad
aa''=a''a,
\quad
a'a''=a''a',
\\
&ba^nb^{-1}=a''^n,
\quad
ba'^nb^{-1}=a^n,
\quad
ba''^nb^{-1}=a'^n,
%\\
%&b^2a^nb^{-2}=a'^n,
%\quad
%b^2a'^nb^{-2}=a''^n,
%\quad
%b^2a''^nb^{-2}=a^n,
%\\
%&ab=ba',
%\quad
%a'b=ba'',
%\quad
%a''b=ba,
%\quad
%ab^2=b^2a'',
%\quad
%a'b^2=b^2a,
%\quad
%a''b^2=b^2a',
\end{split}
\end{eqnarray}
where $e$ denotes the identity.
Group elements can be written by 
$b^ka^\ell a'^m a''^n $ with $\ell,m,n=0,N-1$ and 
$k=0,1,2$. 
When $N/3=$ integer, this group is isomorphic to 
$\Sigma(3N^3)$ \cite{Ishimori:2010au}.
For $N/3 \neq $ integer, this group corresponds to 
$\Delta(3N^2) \times Z_N$.
Here, we use both types of groups.

First, let us study the conjugacy classes of the elements without 
including $b$, 
i.e. $a^{\ell}a'^ma''^n$. 
These elements elements $a^{\ell}a'^ma''^n$ satisfy 
the following relations:
\begin{eqnarray}
ba^{\ell}a'^ma''^nb^{-1}
=a^ma'^na''^\ell,
\quad
b^2a^{\ell}a'^ma''^nb^{-2}
=a^na'^\ell a''^m.
\end{eqnarray}
Note that $ba^\ell a'^\ell a''^\ell b^{-1}=b^2a^\ell a'^\ell a''^\ell b^{-2}=
a^\ell a'^\ell a''^\ell$.
Thus, we find the following conjugacy classes of $a^{\ell}a'^ma''^n$:
\begin{eqnarray}
\begin{split}
&C_1:& &\{ e \},&
\\
&C_1^{(\ell)}:& &\{ a^\ell a'^\ell a''^\ell \},& 
{\rm for}~
\ell=1,\cdots, N-1,
\\
&C_3^{(\ell,m,n)}:& &\{ a^\ell a'^m a''^n,a^m a'^n a''^\ell , a^n a'^\ell a''^m \},&
{\rm for}~
\ell,m,n=0,\cdots, N-1.
\end{split}
\end{eqnarray}
The last classes $C_3^{(\ell,m,n)}$ exclude the case with $\ell=m=n$.
The number of the classes $C_3^{(\ell,m,n)}$ is equal to 
$(N^3-N)/3$, while the number of the classes $C_1^{(\ell)}$ 
is $(N-1)$.

Next, we study the conjugacy classes of the elements including $b$, 
that is 
$ba^\ell a'^m a''^n$ and $b^2a^\ell a'^m a''^n$. 
%There are relations 
%$a^p a'^q a''^r b=ba^ra'^pa''^q$, 
%$a^p a'^q a''^r b^2=b^2a^q a'^r a''^p$. 
%Using them, we have
It is found that 
\begin{eqnarray}
\begin{split}
a^p a'^q a''^r
(ba^\ell a'^m a''^n) a^{-p} a'^{-q} a''^{-r}
&=ba^{\ell +r-p} a'^{m+p-q} a''^{n+q-r},
\\
b a^p a'^q a''^r
(ba^\ell a'^m a''^n) a^{-p} a'^{-q} a''^{-r}b^{-1}
&=b a^{m+p-q} a'^{n+q-r} a''^{\ell+r-p},
\\
b^2 a^p a'^q a''^r
(ba^\ell a'^m a''^n) a^{-p} a'^{-q} a''^{-r}b^{-2}
&=b a^{n+q-r} a'^{\ell+r-p} a''^{m+p-q}.
\end{split}
\end{eqnarray}
Note that the sum of the powers of  $a$, $a'$, and $a''$, 
$\ell + m + n$, does not change.
Then the conjugacy classes of $ba^\ell a'^m a''^n$  are obtained 
as
\begin{eqnarray}
\begin{split}
&C_{N^2}^{(p)}:& &\{ ba^\ell a'^m a''^{p-\ell-m}|
\ell,m=0,\cdots,N-1 \},
\quad
{\rm for}~
p=0,\cdots,N-1.&
\end{split}
\end{eqnarray}
Similarly, we obtain the conjugacy classes of $b^2a^\ell a'^m a''^n$ 
as   
\begin{eqnarray}
\begin{split}
&C'^{(p)}_{N^2}:& &\{ b^2a^\ell a'^m a''^{p'-\ell-m}|
\ell,m=0,\cdots,N-1 \},
\quad
{\rm for}~
p=0,\cdots,N-1.&
&
\end{split}
\end{eqnarray}
Then, the total number of conjugacy classes is
\begin{eqnarray}
1+(N-1)+(N^3-N)/3+N+N=\frac13 N(N^2+8).
\end{eqnarray}

Now, let us study representations of our group.
Suppose that there are $m_n$ $n$-dimensional irreducible 
representations, where group elements are represented by 
$(n \times n)$ matrices.
The identity $e$ is always represented by the 
$(n \times n)$ identity matrix and its character 
$\chi(e)$ is obtained as $\chi(e)=n$ on the $n$-dimensional 
representation.
The orthogonality condition of characters requires
$\sum_n n^2 m_n $ to be equal to the order of the group,\footnote{
See e.g. \cite{Ishimori:2010au} and references therein.} i.e.,
\begin{eqnarray}
\label{eq:group-1}
m_1+4m_2+9m_3+\cdots =3N^3.
\end{eqnarray}
In addition, the total number of irreducible 
representations should be equal to the total number 
of the conjugacy classes, i.e.,
\begin{eqnarray}
\label{eq:group-2}
m_1+m_2+m_3+\cdots=\frac13 N(N^2+8).
\end{eqnarray}
The solution of Eqs.~(\ref{eq:group-1}) and (\ref{eq:group-2})  is 
obtained as $(m_1,m_2,m_3)=(3N,0,N(N^2-1)/3)$. 
That implies that there are $3N$ singlets and $N(N^2-1)/3$ triplets.

%\subsection{Representation and multiplication}

It is straightforward to derive the $3N$ singlet representations, 
${\bf 1}^\ell_k$ with $k = 0, \cdots, N-1$ and $\ell =0,1,2$.
Their characters for $a$, $a'$, $a''$ and $b$ are obtained as
$\chi_{1_{k}^\ell}(a)=\chi_{1_{k}^\ell}(a')=\chi_{1_{k}^\ell}(a'')=\rho^k$  
and $\chi_{1_{k}^\ell}(b)=\omega^\ell$, where $\rho=e^{2\pi i/N}$, 
and $\omega=e^{2\pi i/3}$.
These are shown in Table \ref{tab:character}, 
where $h$ denotes the order of the element $g$ in the conjugacy class,
i.e., $g^h=e$.
Note that these characters are nothing but representations for 
 $a$, $a'$, $a''$ and $b$ on ${\bf 1}^\ell_k$.

On the other hand, we can write $a$, $a'$, $a''$, and $b$ as 
%The number of triplets is $N(N^2-1)/3$. 
%We write triplets by ${\bf 3}_{[\ell][m][n]}$ whose 
%representations for $a$, $a'$, $a''$, and $b$ are written by
\begin{eqnarray}
a=
\begin{pmatrix}
\rho^\ell &0&0\\
0&\rho^m&0\\
0&0&\rho^n
\end{pmatrix},
\quad
a'=
\begin{pmatrix}
\rho^n &0&0\\
0&\rho^\ell&0\\
0&0&\rho^m
\end{pmatrix},
\quad
a''=
\begin{pmatrix}
\rho^m &0&0\\
0&\rho^n&0\\
0&0&\rho^\ell
\end{pmatrix},
\quad
b=
\begin{pmatrix}
0&1&0\\
0&0&1\\
1&0&0
\end{pmatrix},
\end{eqnarray}
on the triplet ${\bf 3}_{[\ell][m][n]}$, where 
the notation $[\ell][m][n]$ means
\begin{eqnarray}
[\ell][m][n]
=(\ell,m,n), ~(m,n,\ell), ~\rm{or}~(n,\ell,m).
\end{eqnarray}
Their characters are shown Table \ref{tab:character}.

We write components of triplets as
\begin{eqnarray}
{\bf 3}_{[\ell][m][n]}
=\begin{pmatrix}
x_\ell \\
x_m \\
x_n
\end{pmatrix}.
\end{eqnarray}
Subscripts of the components describe $Z_N$ charges. 
The tensor product between two triplets is given by
\begin{eqnarray}
\begin{split}
&\begin{pmatrix}
x_\ell \\
x_m \\
x_n
\end{pmatrix}_{{\bf 3}_{[\ell][m][n]}}
\times
\begin{pmatrix}
y_{\ell'} \\
y_{m'} \\
y_{n'}
\end{pmatrix}_{{\bf 3}_{[\ell'][m'][n']}}
\\
&=\begin{pmatrix}
x_\ell y_{\ell'} \\
x_m y_{m'} \\
x_n y_{n'}
\end{pmatrix}_{{\bf 3}_{[\ell+\ell'][m+m'][n+n']}}
+\begin{pmatrix}
x_m y_{n'} \\
x_n y_{\ell'} \\
x_\ell  y_{m'}
\end{pmatrix}_{{\bf 3}_{[m+n'][n+\ell'][\ell+m']}}
+\begin{pmatrix}
x_n y_{m'} \\
x_\ell y_{n'} \\
x_m y_{\ell'}
\end{pmatrix}_{{\bf 3}_{[n+m'][\ell+n'][m+\ell']}}.
\end{split}
\end{eqnarray}
If all the subscripts of ${\bf 3}_{[\ell+\ell'][m+m'][n+n']}$, 
${\bf 3}_{[m+n'][n+\ell'][\ell+m']}$ or ${\bf 3}_{[n+m'][\ell+n'][m+\ell']}$
are the same, 
such a triplet can be decomposed into singlets 
as 
\begin{eqnarray}
(x_a,x_b,x_c)_{{\bf 3}_{[k][k][k]}}
=(x_a+x_b+x_c)_{{\bf 1}_{k}^0}
+(x_a+\omega^2 x_b+\omega x_c)_{{\bf 1}_{k}^1}
+(x_a+\omega x_b+\omega^2 x_c)_{{\bf 1}_{k}^2}. 
\end{eqnarray}
The tensor product between singlet and triplet is obtained as 
\begin{eqnarray}
\begin{pmatrix}
x_\ell \\
x_m \\
x_n
\end{pmatrix}_{{\bf 3}_{[\ell][m][n]}}
\times
(y)_{{\bf 1}_k^{k'}}
=\begin{pmatrix}
x_\ell y\\
\omega^{k'}x_my \\
\omega^{2k'}x_ny
\end{pmatrix}_{{\bf 3}_{[\ell+k][m+k][n+k]}}.
\end{eqnarray}
The tensor product between singlets is obtained as 
\begin{eqnarray}
(x)_{{\bf 1}_k^\ell}
\times
(y)_{{\bf 1}_{k'}^{\ell'}}
=(xy)_{{\bf 1}_{k+k'}^{\ell+\ell'}}.
\end{eqnarray}

\begin{table}[t]
\begin{center}
\begin{tabular}{|c|c|c|c|}
\hline
   & $h$ & $\chi_{1^{\ell}_{m}}$ & $\chi_{3_{[\ell][m][n]}}$  
\\ \hline
   %%%%%%%%%%%%%%%%%%%%%%
$C_1$ & 1 &  1 &3    \\ \hline
$C_1^{(p)}$ & $\frac{N}{\gcd(N,p)}$ &  $\rho^{pm}$ & $\rho^{p(\ell+m+n)}$    \\ \hline
$C_3^{(p,q,r)}$ & $\frac{N}{\gcd(N,p,q,r)}$ &  $\rho^{(p+q+r)m}$ 
& $\rho^{p\ell+qn+mr}+\rho^{pm+q\ell+rn}+\rho^{pn+qm+r\ell}$    \\ \hline
$C_{N^2}^{(p)}$ & $\frac{3N}{\gcd(N,p)}$ &  $\omega^\ell \rho^{pm}$ & $0$    \\ \hline
$C_{N^2}'^{(p)}$ & $\frac{3N}{\gcd(N,p)}$ &  $\omega^{2\ell} \rho^{pm}$ & $0$    \\ \hline
\end{tabular}
\end{center}
\caption{Characters.}
\label{tab:character}
\end{table}

\section{Lepton mass and mixing}

We use the group $(Z_N \times Z_N \times Z_N) \rtimes Z_3$ 
as the lepton flavor symmetry.
We consider the case with $N/4=$ integer.
We also assume the additional $Z_4$ flavor symmetry.
We assign the quantum numbers of the three families 
of 
lepton doublets $(\ell_e,\ell_\mu,\ell_\tau)$ and 
lepton singlets $(e^c,\mu^c,\tau^c)$ and the Higgs fields 
$H_{u,d}$ under the flavor symmetry as shown in Table \ref{tab:model}.
These denote chiral superfields.
In addition, we introduce the flavon fields, 
$\chi^\nu_0$, $(\chi^\nu_1,\chi^\nu_2,\chi^\nu_3)$  and 
$(\chi^\ell_1,\chi^\ell_2,\chi^\ell_3)$, which also denote 
chiral superfields.
Their quantum numbers under $(Z_N \times Z_N \times Z_N) \rtimes Z_3$  
and $Z_4$ are also shown in Table \ref{tab:model}.

\begin{table}[h]
\begin{tabular}{|c|cccccccc|}
\hline
&$e^c$&$\mu^c$&$\tau^c$&$(\ell_e,\ell_\mu,\ell_\tau)$
&$H_{u,d}$&$\chi^\nu_0$&$(\chi^\nu_1,\chi^\nu_2,\chi^\nu_3)$ 
&$(\chi^\ell_1,\chi^\ell_2,\chi^\ell_3)$
\\ \hline
$(Z_N^3) \rtimes Z_3$  & ${\bf 1}^1_{3N/4+p}$& ${\bf 1}^2_{3N/4+q}$& ${\bf 1}^0_{3N/4}$
& ${\bf 3}_{[3N/4][N/4][N/4]}$
& ${\bf 1}^0_{0}$& ${\bf 1}^0_{N/2}$& ${\bf 3}_{[N/2][0][0]}$& ${\bf 3}_{[N-1][0][0]}$
\\
$Z_4$ & $1$& $1$& $1$
& $3$& $0$
& $2$& $2$& $0$
\\
\hline
\end{tabular}
\caption{Quantum numbers of leptons, Higgs fields and flavon fields 
under $(Z_N \times Z_N \times Z_N) \rtimes Z_3$  and $Z_4$. 
Parameters $p$ and $q$ of 
$(Z_N \times Z_N \times Z_N) \rtimes Z_3$ charges 
are arbitrary integers.}
\label{tab:model}
\end{table}

The terms relevant to charged lepton masses in the superpotential are 
obtained as 
\begin{eqnarray}
\begin{split}
w_e
=&y^e e^c(\ell_e(\chi_1^\ell)^{N/2}
+\ell_\mu(\chi_2^\ell)^{N/2}
+\ell_\tau(\chi_3^\ell)^{N/2})H_d
(\chi_1^\ell\chi_2^\ell\chi_3^\ell)^p
/\Lambda^{N/2+3p}
\\
&+y^\mu \mu^c
(\ell_e(\chi_1^\ell)^{N/2}
+\ell_\mu(\chi_2^\ell)^{N/2}
+\ell_\tau(\chi_3^\ell)^{N/2})H_d
(\chi_1^\ell\chi_2^\ell\chi_3^\ell)^q
/\Lambda^{N/2+3q}
\\
&+y^\tau \tau^c
(\ell_e(\chi_1^\ell)^{N/2}
+\ell_\mu(\chi_2^\ell)^{N/2}
+\ell_\tau(\chi_3^\ell)^{N/2})H_d/\Lambda^{N/2},
\end{split}
\end{eqnarray}
%The charges $a$ and $b$ are given with 
%the dependence of $N$. 
where $\Lambda$ is the cut-off scale and 
$y^e, y^\mu, y^\tau$ are dimensionless couplings.
Similarly, the terms relevant to neutrino masses in the 
superpotential are obtained as 
\begin{eqnarray}
\begin{split}
w_\nu
=&y_{1}^\nu(\ell_e\ell_e+\ell_\mu\ell_\mu+\ell_\tau\ell_\tau)
H_uH_u \chi_0^\nu/\Lambda^2
\\
&+y_{2}^\nu 
(\ell_e\ell_\mu\chi_3^\nu+\ell_e\ell_\tau\chi_2^\nu
+\ell_\mu\ell_\tau\chi_1^\nu)H_uH_u/\Lambda^2,
\end{split}
\end{eqnarray}
where $y^\nu_{1,2}$ are dimensionless couplings.

We assume that all scalar fields develop VEVs 
which are written by
\begin{eqnarray}
\langle H_{u,d}\rangle
=v_{u,d},
\quad
\langle \chi_{0}^\nu \rangle
=\alpha_\nu \Lambda,
\quad
\langle (\chi_{1}^\nu,\chi_2^\nu,\chi_3^\nu)\rangle
=(\alpha_1^\nu,\alpha_2^\nu,\alpha_3^\nu)\Lambda,
\quad
\langle (\chi_{1}^\ell,\chi_2^\ell,\chi_3^\ell)\rangle
=(\alpha_1^\ell,\alpha_2^\ell,\alpha_3^\ell)\Lambda.
\end{eqnarray}
Furthermore, as the vacuum alignment, we assume
\begin{eqnarray}
\langle (\chi_{1}^\nu,\chi_2^\nu,\chi_3^\nu)\rangle
=\alpha_\nu'(1,0,0)\Lambda,
\quad
\langle (\chi_{1}^\ell,\chi_2^\ell,\chi_3^\ell)\rangle
=\alpha_\ell(1,\rho,\rho')\Lambda,
\end{eqnarray}
where $\rho$ and $\rho'$ are 
phases of $Z_N$ so that we write 
$\rho=e^{2\pi i m/N}$, $\rho'=e^{2\pi i n/N}$ with 
integers $m$ and $n$.

Taking the vacuum alignment, 
mass matrices of charged leptons and neutrinos, $M_e$ and $M_\nu$, 
are given by
\begin{eqnarray}
\begin{split}
&M_e
=
v_d\alpha_\ell^{N/2}
\begin{pmatrix}
y^e\alpha_\ell^{3p}(\rho\rho')^p&0&0\\
0&y^\mu \alpha_\ell^{3q}(\rho\rho')^q&0\\
0&0&y^\tau 
\end{pmatrix}
\begin{pmatrix}
1 &\omega^2 &\omega  \\
1 &\omega  &\omega^2 \\
1 &1 &1 \\
\end{pmatrix}
\begin{pmatrix}
1&0&0\\
0&\rho&0\\
0&0&\rho'
\end{pmatrix},
\\
&M_\nu
=
\frac{v_u^2}{\Lambda}
\begin{pmatrix}
y_1^\nu \alpha_\nu &0 &0\\
0&y_1^\nu \alpha_\nu &y_2^\nu \alpha_\nu'\\
0&y_2^\nu \alpha_\nu' &y_1^\nu \alpha_\nu
\end{pmatrix} .
\end{split}
\end{eqnarray}
Except additional phases $\rho$ and $\rho'$, 
the mass matrices are the same as those of 
the $A_4$ model \cite{Altarelli:2005yp}. 
They can be easily diagonalized. 
The masses of charged lepton are
\begin{eqnarray}
m_e=y^e\alpha_\ell^{N/2+3p} v_d,
\quad
m_\mu=y^\mu \alpha_\ell^{N/2+3q} v_d,
\quad
m_\mu=y^\tau\alpha_\ell^{N/2} v_d.
\end{eqnarray}
Taking $p\approx 2q$, 
we can realize the mass hierarchy of charged leptons 
when the Yukawa couplings, $y^e, y^\mu$ and $y^\tau$, 
are of the same order each other.
Suppose that Yukawa couplings are of ${\cal O}(1)$.
Then, the value of $3p$ should 
be taken as the same magnitude of $N$. 

The mixing matrix becomes
\begin{eqnarray}
\begin{split}
U_{\rm MNS}
&=\frac{1}{\sqrt3}\begin{pmatrix}
1 &\omega &\omega^2  \\
1 &\omega^2  &\omega \\
1 &1 &1 \\
\end{pmatrix}
\begin{pmatrix}
1&0&0\\
0&\rho^{-1}&0\\
0&0&\rho'^{-1}
\end{pmatrix}
\begin{pmatrix}
1&0&0\\
0&1/\sqrt2 &-1/\sqrt2\\
0&1/\sqrt2&1/\sqrt2
\end{pmatrix}
\\
&=\frac{1}{\sqrt3}
\begin{pmatrix}
1 &\omega^2(\rho+\omega^2 \rho')/\sqrt2\rho\rho'  
&\omega^2 (\rho-\omega^2 \rho')/\sqrt2\rho\rho' \\
1 &\omega (\rho+\omega \rho')/\sqrt2\rho\rho' 
&\omega (\rho-\omega \rho')/\sqrt2\rho\rho'  \\
1 &(\rho+\rho')/\sqrt2\rho\rho' &(\rho-\rho')/\sqrt2\rho\rho' \\
\end{pmatrix}.
\end{split}
\end{eqnarray}
Assuming $\rho\approx \omega^2\rho'$, 
the $(1,3)$ element can be small enough to be fitted to experiments. 
Then we obtain 
\begin{eqnarray}
|m_1|=|y_1^\nu \alpha_\nu+y_2^\nu \alpha_\nu'|\frac{v_u^2}{\Lambda},
\quad
|m_2|=|y_1^\nu \alpha_\nu| \frac{v_u^2}{\Lambda},
\quad
|m_3|=|y_1^\nu \alpha_\nu-y_2^\nu \alpha_\nu'|\frac{v_u^2}{\Lambda}.
\end{eqnarray}
To explain the neutrino mass difference, we need
\begin{eqnarray}
2|y_1^\nu\alpha_\nu| |y_2^\nu \alpha_\nu'|
(\cos \Delta) \frac{v_u^4}{\Lambda^2}
=-\frac12 \Delta m_{\rm atm}^2,
\quad
|y_2^\nu\alpha_\nu'|^2\frac{v_u^4}{\Lambda^2}
=\frac12 \Delta m_{\rm atm}^2-\Delta m_{\rm sol}^2,
\end{eqnarray}
where $\Delta$ is the phase difference between 
$y_1^\nu\alpha_\nu$ and $y_2^\nu\alpha_\nu'$. 
Since experiments show 
$|\Delta m_{\rm atm}^2|/2>\Delta m_{\rm sol}^2$, 
the second equation gives $\Delta m_{\rm atm}^2>0$, 
i,e, the normal mass hierarchy. 
Using them, we obtain the neutrino mass spectrum as
\begin{eqnarray}
\begin{split}
&|m_1|^2=\frac{\Delta m_{\rm atm}^2}
{8(1-2r)\cos^2 \Delta}
-\Delta m_{\rm sol}^2,
\quad
|m_2|^2=\frac{\Delta m_{\rm atm}^2}
{8(1-2r)\cos^2 \Delta},
\\
&|m_3|^2=\frac{\Delta m_{\rm atm}^2}
{8(1-2r)\cos^2 \Delta}
+\Delta m_{\rm atm}^2-\Delta m_{\rm sol}^2,
\end{split}
\end{eqnarray}
where $r=\Delta m_{\rm sol}^2/\Delta m_{\rm atm}^2$. 
Inserting the best-fit values for $\Delta m_{\rm atm}^2$ and 
$\Delta m_{\rm atm}^2$ and $\cos^2\Delta=1$, 
we obtain $|m_1|=16$meV, $|m_2|=18$meV, and $|m_3|=52$meV.
The smallest value of sum of neutrino masses is
$86$meV.

In the usual notation, mixing angles can be expressed by
\begin{eqnarray}
\begin{split}
\sin^2\theta_{13}
&=\frac{2+\cos(2\pi(m-n)/N)+\sqrt3\sin(2\pi(m-n)/N)}{6},
\\
\sin^2\theta_{12}
&=\frac{2}
{4-\cos(2\pi(m-n)/N)-\sqrt3\sin(2\pi(m-n)/N)},
\\
\sin^2\theta_{23}
&=1-
\frac{4\sin^2(\pi(m-n)/N)}
{4-\cos(2\pi(m-n)/N)-\sqrt3\sin(2\pi(m-n)/N)}.
\end{split}
\end{eqnarray}
The Dirac CP phase is always vanishing. 
When $N=8$, the smallest $|U_{e3}|$ is obtained as
\begin{eqnarray}
|U_{e3}|=  \sqrt{(2-\sqrt{2+\sqrt3})/6},
\end{eqnarray}
which is about $0.107$. 
In this case, we find\footnote{These mixing pattern 
is obtained by another approach\cite{deAdelhartToorop:2011re,Toorop:2011jn}}
\begin{eqnarray}
\sin\theta_{12}=\frac{2}{\sqrt{8+\sqrt2+\sqrt6}}, \qquad 
\sin\theta_{23}
=\frac{\sqrt{5-3\sqrt2-\sqrt3+\sqrt6}}{\sqrt{7-3\sqrt2-\sqrt3+\sqrt6}}.
\end{eqnarray} 
For $N=12$, $U_{e3}$ can be vanishing and mixing matrix becomes 
tri-bimaximal mixing.

\begin{figure}[ttb]
\includegraphics[width=7.0cm]{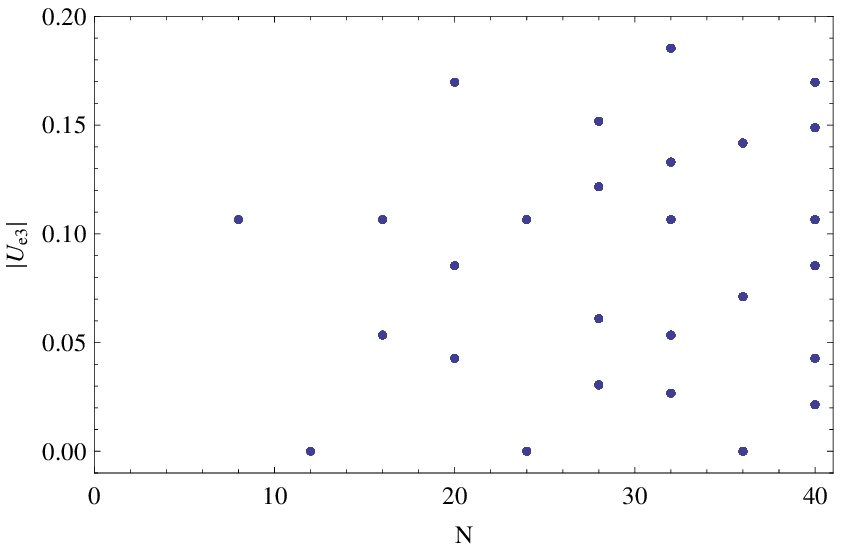}
\quad
\includegraphics[width=7.0cm]{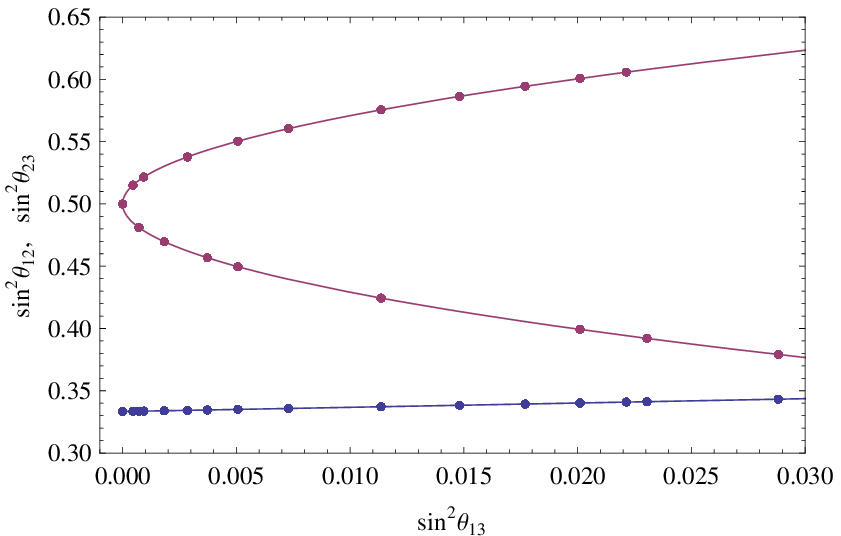}
\caption{The left figure shows the relation between 
$N$ and $|U_{e3}|$ for possible values of $\rho$ and $\rho'$. 
The right figure shows the values of $\sin^2\theta_{12}$, 
$\sin^2\theta_{13}$, and $\sin^2\theta_{23}$ for the region 
$0\leq N\leq 40$ in the case $N/4$ is integer. Continuous line 
is the correlation of mixing angles in the limit of $N\rightarrow \infty$. }
\end{figure}

We have calculated possible mixing angles which can be 
consistent in experiments up to $N=40$ in Fig. 1. 
By fixing $\rho$ and $\rho'$, we can predict 
mixing angles. In Fig. 2, we estimate the effective mass 
of double beta decay 
$m_{ee}=|m_1 U_{e1}^2+m_2 U_{e2}^2+m_3 U_{e3}^2|$. 
Due to Majorana phases and uncertainty of neutrino mass differences, 
the effective mass is predicted in continuous line with some fluctuation. 
For $N=8$, the minimum value of the mass of double beta decay 
is predicted about $5.3$ meV,  and for other $N$ it is about $4.6$ meV. 

\begin{figure}[ttb]
\includegraphics[width=7.0cm]{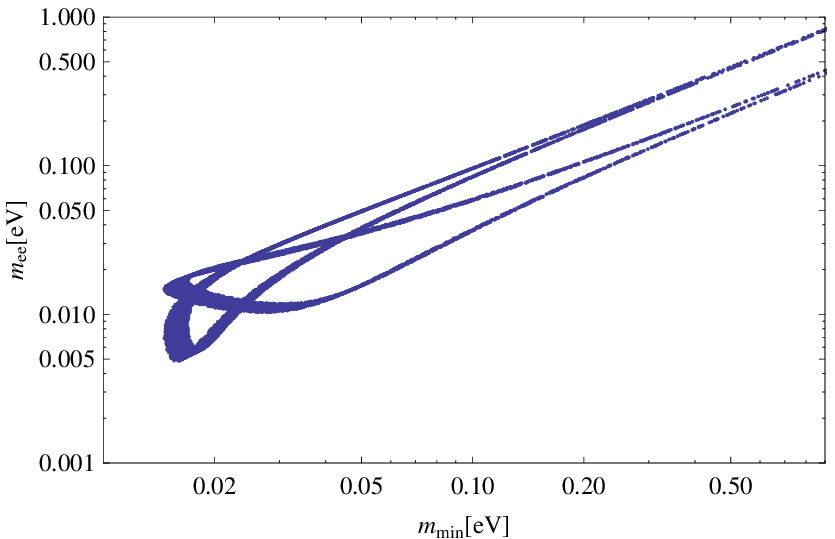}
\quad
\includegraphics[width=7.0cm]{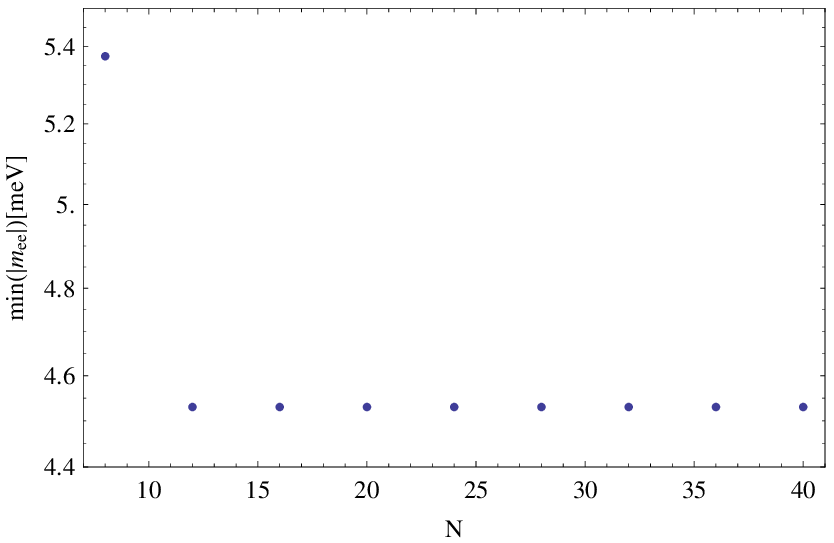}
\caption{The left figure shows the relation between 
minimum neutrino mass and effective mass of double beta decay 
for $N=8$. 
The right figure shows the minimum value of 
the effective mass of double beta decay for each $N$.}
\end{figure}

\section{Vacuum alignment}

Here, we discuss the vacuum alignment of flavon VEVs.
We study it in two ways, 
one is to use the potential analysis in four-dimensional 
field theory and the other is to use 
boundary conditions in extra dimensions 
\cite{Kobayashi:2008ih,Seidl:2008yf,Adulpravitchai:2010na,Burrows:2010wz}.

\subsection{Potential analysis}
Here, we study the vacuum alignment of flavon VEVs 
by using four-dimensional potential analysis.
The superpotential terms of flavon fields $\chi^\nu_i$ ($i=0,1,2,3$) up to 
four dimensional operators are
\begin{eqnarray}
\begin{split}
w_{f}
=&M_1(\chi_0^\nu)^2
+M_2((\chi_1^\nu)^2+(\chi_2^\nu)^2+(\chi_3^\nu)^2)
+\lambda_1(\chi_0^\nu)^4
+\lambda_2(\chi_0^\nu)^2((\chi_1^\nu)^2+(\chi_2^\nu)^2+(\chi_3^\nu)^2)
\\
&+\lambda_3((\chi_1^\nu)^2+\omega(\chi_2^\nu)^2+\omega^2(\chi_3^\nu)^2)
((\chi_1^\nu)^2+\omega^2(\chi_2^\nu)^2+\omega(\chi_3^\nu)^2)
\\
&+\lambda_4((\chi_1^\nu)^2+(\chi_2^\nu)^2+(\chi_3^\nu)^2)
((\chi_1^\nu)^2+(\chi_2^\nu)^2+(\chi_3^\nu)^2)
\\
&+\lambda_5((\chi_2^\nu\chi_3^\nu)^2
+(\chi_3^\nu\chi_1^\nu)^2+(\chi_1^\nu\chi_2^\nu)^2)
+\lambda_6((\chi_1^\nu)^4
+(\chi_2^\nu)^4+(\chi_3^\nu)^4),
\end{split}
\end{eqnarray}
where $\lambda_i={\bar\lambda}_i/\Lambda$ 
($i=1,2,\cdots,6$) with dimensionless 
coupling constants ${\bar\lambda}_i$. 
The conditions of the potential minimum are written as 
\begin{eqnarray}
\begin{split}
2M_1\chi_0^\nu
+4\lambda_1(\chi_0^\nu)^3
+2\lambda_2(\chi_0^\nu)
((\chi_1^\nu)^2+(\chi_2^\nu)^2+(\chi_3^\nu)^2)
&=0,
\\
2M_2\chi_1^\nu
+2\lambda_2(\chi_0^\nu)^2\chi_1^\nu
+2\lambda_3\chi_1^\nu
(2(\chi_1^\nu)^2-(\chi_2^\nu)^2-(\chi_3^\nu)^2)
&\\
+4\lambda_4\chi_1^\nu
((\chi_1^\nu)^2+(\chi_2^\nu)^2+(\chi_3^\nu)^2)
+2\lambda_5 \chi_1^\nu((\chi_2^\nu)^2+(\chi_3^\nu)^2)
+4\lambda_6 (\chi_1^\nu)^3
&=0,
\\
2M_2\chi_2^\nu
+2\lambda_2(\chi_0^\nu)^2\chi_2^\nu
+2\lambda_3\chi_2^\nu
(-(\chi_1^\nu)^2+2(\chi_2^\nu)^2-(\chi_3^\nu)^2)
&\\
+4\lambda_4\chi_2^\nu
((\chi_1^\nu)^2+(\chi_2^\nu)^2+(\chi_3^\nu)^2)
+2\lambda_5 \chi_2^\nu((\chi_1^\nu)^2+(\chi_3^\nu)^2)
+4\lambda_6 (\chi_2^\nu)^3
&=0,
\\
2M_2\chi_3^\nu
+2\lambda_2(\chi_0^\nu)^2\chi_3^\nu
+2\lambda_3\chi_3^\nu
(-(\chi_1^\nu)^2-(\chi_2^\nu)^2+2(\chi_3^\nu)^2)
&\\
+4\lambda_4\chi_3^\nu
((\chi_1^\nu)^2+(\chi_2^\nu)^2+(\chi_3^\nu)^2)
+2\lambda_5 \chi_3^\nu((\chi_1^\nu)^2+(\chi_2^\nu)^2)
+4\lambda_6 (\chi_3^\nu)^3
&=0.
\end{split}
\end{eqnarray}
There is a solution we have used as the vacuum alignment
\begin{eqnarray}
\begin{split}
&\chi_0^\nu
=\sqrt{\frac{\lambda_2M_2-2(\lambda_3+\lambda_4+\lambda_6)M_1}
{4\lambda_1(\lambda_3+\lambda_4+\lambda_6)-\lambda_2^2}},
\\
&\chi_1^\nu
=\sqrt{\frac{\lambda_2M_1-2\lambda_1M_2}
{4\lambda_1(\lambda_3+\lambda_4)-\lambda_2^2}},
\quad
\chi_2^\nu
=0,
\quad
\chi_3^\nu
=0.
\end{split}
\end{eqnarray}

The leading and next leading order terms of $\chi^\ell_i$ 
($i=1,2,3$) in the superpotential are obtained as
\begin{eqnarray}
\begin{split}
w_{f'}
=&m'((\chi_1^\ell)^N+(\chi_2^\ell)^N+(\chi_3^\ell)^N)
\\
&+\lambda'_1((\chi_1^\ell)^N+\omega(\chi_2^\ell)^N
+\omega^2(\chi_3^\ell)^N)
((\chi_1^\ell)^N+\omega^2(\chi_2^\ell)^N
+\omega(\chi_3^\ell)^N)
\\
&+\lambda'_2((\chi_1^\ell)^N+(\chi_2^\ell)^N+(\chi_3^\ell)^N)
((\chi_1^\ell)^N+(\chi_2^\ell)^N+(\chi_3^\ell)^N)
\\
&+\lambda'_3((\chi_2^\ell\chi_3^\ell)^N+(\chi_1^\ell\chi_3^\ell)^N
+(\chi_1^\ell\chi_2^\ell)^N)
+\lambda'_4((\chi_1^\ell)^{2N}+(\chi_2^\ell)^{2N}
+(\chi_3^\ell)^{2N}),
\end{split}
\end{eqnarray}
where $m'={\bar m}'/\Lambda^{N-3}$ and 
$\lambda_i'={\bar\lambda}'_i/\Lambda^{2N-3}$ 
($i=1,2,3,4$) with 
dimensionless constants ${\bar m}'$ 
and ${\bar\lambda}_i'$. 
%The conditions of the potential minimum are written as 
%\begin{eqnarray}
%\begin{split}
%Nm'(\chi_1^\ell)^{N-1}
%+N\lambda_1'(\chi_1^\ell)^{N-1}
%(2(\chi_1^\ell)^{N}-(\chi_2^\ell)^N-(\chi_3^\ell)^N)
%+2N\lambda_2'(\chi_1^\ell)^{N-1}
%((\chi_1^\ell)^{N}+(\chi_2^\ell)^N+(\chi_3^\ell)^N)
%&\\
%+N\lambda_3'(\chi_1^\ell)^{N-1}
%((\chi_2^\ell)^N+(\chi_3^\ell)^N)
%+2N\lambda_4'(\chi_1^\ell)^{2N-1}
%=0,&
%\\
%Nm'(\chi_2^\ell)^{N-1}
%+N\lambda_1'(\chi_2^\ell)^{N-1}
%(-(\chi_1^\ell)^{N}+2(\chi_2^\ell)^N-(\chi_3^\ell)^N)
%+2N\lambda_2'(\chi_2^\ell)^{N-1}
%((\chi_1^\ell)^{N}+(\chi_2^\ell)^N+(\chi_3^\ell)^N)
%&\\
%+N\lambda_3'(\chi_2^\ell)^{N-1}
%((\chi_3^\ell)^N+(\chi_1^\ell)^N)
%+2N\lambda_4'(\chi_2^\ell)^{2N-1}
%=0,&
%\\
%Nm'(\chi_3^\ell)^{N-1}
%+N\lambda_1'(\chi_3^\ell)^{N-1}
%(-(\chi_1^\ell)^{N}-(\chi_2^\ell)^N+2(\chi_3^\ell)^N)
%+2N\lambda_2'(\chi_3^\ell)^{N-1}
%((\chi_1^\ell)^{N}+(\chi_2^\ell)^N+(\chi_3^\ell)^N)
%&\\
%%+N\lambda_3'(\chi_3^\ell)^{N-1}
%((\chi_1^\ell)^N+(\chi_2^\ell)^N)
%+2N\lambda_4'(\chi_3^\ell)^{2N-1}
%=0,&
%\end{split}
%\end{eqnarray}
When $\chi_1^\ell\neq 0$, $\chi_2^\ell\neq 0$, 
and $\chi_3^\ell\neq 0$, we have
\begin{eqnarray}
\begin{split}
m'+\lambda_1'
(2(\chi_1^\ell)^{N}-(\chi_2^\ell)^N-(\chi_3^\ell)^N)
+2\lambda_2'
((\chi_1^\ell)^{N}+(\chi_2^\ell)^N+(\chi_3^\ell)^N)
&\\
+\lambda_3'
((\chi_2^\ell)^N+(\chi_3^\ell)^N)
+2\lambda_4'(\chi_1^\ell)^{N}
=0,&
\\
m'+\lambda_1'
(-(\chi_1^\ell)^{N}+2(\chi_2^\ell)^N-(\chi_3^\ell)^N)
+2\lambda_2'
((\chi_1^\ell)^{N}+(\chi_2^\ell)^N+(\chi_3^\ell)^N)
&\\
+\lambda_3'
((\chi_3^\ell)^N+(\chi_1^\ell)^N)
+2\lambda_4'(\chi_2^\ell)^{N}
=0,&
\\
m'+\lambda_1'
(-(\chi_1^\ell)^{N}-(\chi_2^\ell)^N+2(\chi_3^\ell)^N)
+2\lambda_2'
((\chi_1^\ell)^{N}+(\chi_2^\ell)^N+(\chi_3^\ell)^N)
&\\
+\lambda_3'
((\chi_1^\ell)^N+(\chi_2^\ell)^N)
+2\lambda_4'(\chi_3^\ell)^{N}
=0.&
\end{split}
\end{eqnarray}
Moreover, when $3\lambda_1'-\lambda_3'+2\lambda_4'\neq 0$, 
we have the relations
\begin{eqnarray}
(\chi_1^\ell)^N=(\chi_2^\ell)^N=(\chi_3^\ell)^N
=-\frac{m'}{6\lambda_2'+2\lambda_3'+2\lambda_4'}.
\end{eqnarray}
Then we can realize the vacuum alignment
$\chi_2=\chi_1 \rho$ and  $\chi_3=\chi_1 \rho'$. 

There are the cross terms between $\chi^\nu$ and $\chi^\ell$. 
For instance, if 
$(\chi^\ell_i/\Lambda)^{N/2}\sim \chi^\nu_0\sim\chi^\nu_1$, 
the most effective terms are
\begin{eqnarray}
\begin{split}
\Delta w_f
=&(\lambda''_1 (\chi_0^\nu)^2
+\lambda''_2 ((\chi_1^\nu)^2+(\chi_2^\nu)^2+(\chi_3^\nu)^2) )
((\chi_1^\ell)^N+(\chi_2^\ell)^N+(\chi_3^\ell)^N))
\\
&+\lambda''_3 \chi_0^\nu 
(\chi_1^\nu(\chi_2^\ell\chi_3^\ell)^{N/2}
+\chi_2^\nu(\chi_1^\ell\chi_3^\ell)^{N/2}
+\chi_3^\nu(\chi_1^\ell\chi_2^\ell)^{N/2})
\\
&+\lambda''_4 
(\chi_2^\nu\chi_3^\nu(\chi_2^\ell\chi_3^\ell)^{N/2}
+\chi_1^\nu\chi_3^\nu(\chi_1^\ell\chi_3^\ell)^{N/2}
+\chi_1^\nu\chi_2^\nu(\chi_1^\ell\chi_2^\ell)^{N/2}).
\end{split}
\end{eqnarray}
These terms would violate the vacuum alignment that we obtain above. 
We need to assume these cross terms should be suppressed sufficiently 
in this system. 

\subsection{Vacuum alignment on orbifold}

Here, we discuss another way to realize the vacuum alignment 
for $(\chi_1^\nu,\chi_2^\nu,\chi_3^\nu )$ and 
$(\chi_1^\ell,\chi_2^\ell,\chi_3^\ell )$, 
using extra dimensional field theory.
We study the flavor symmetry breaking by boundary conditions 
on the orbifold
\cite{Kobayashi:2008ih,Seidl:2008yf,Adulpravitchai:2010na,
Burrows:2010wz}.\footnote{
The orbifold models are also interesting as the origin of 
non-Abelian discrete flavor symmetries 
\cite{Altarelli:2006kg,Adulpravitchai:2009id,
Kobayashi:2004ya,Kobayashi:2006wq,Ko:2007dz,Abe:2010iv}.}

We consider  eight-dimensional field theory on 
the $T^2/Z_2\times T^2/Z_3$ orbifold. 
The fifth and sixth dimensions, $(x^5,x^6)$, are 
compactified on the $T^2/Z_2$ orbifold and 
the seventh and eight dimensions, $(x^7,x^8)$, are 
compactified on the $T^2/Z_3$ orbifold.
It is useful to denote the extra-dimensional coordinates 
$(x^5,x^6,x^7,x^8)$ by a complex space $z=x^5+i x^6$ and 
$z'=x^7+i x^8$. 
We assume that $(\chi_1^\nu,\chi_2^\nu,\chi_3^\nu)$ and 
$(\chi_1^\ell,\chi_2^\ell,\chi_3^\ell)$ live on 
$T^2/Z_2$ and $T^2/Z_3$, respectively.
We have to fix the boundary conditions of these fields 
under the $Z_2$ twist $P$ and the $Z_3$ twist $R$, i.e. 
\begin{eqnarray}
\begin{pmatrix}
\chi_1^\nu(-z) \\
\chi_2^\nu(-z) \\
\chi_3^\nu (-z)
\end{pmatrix}
= P
\begin{pmatrix}
\chi_1^\nu(z) \\
\chi_2^\nu(z) \\
\chi_3^\nu (z)
\end{pmatrix},\qquad 
\begin{pmatrix}
\chi_1^\ell(\omega z') \\
\chi_2^\ell(\omega z') \\
\chi_3^\ell (\omega z')
\end{pmatrix}
= P
\begin{pmatrix}
\chi_1^\ell(z') \\
\chi_2^\ell(z') \\
\chi_3^\ell (z')
\end{pmatrix},
\end{eqnarray}
where $\omega = e^{2\pi i/3}$.

When  we take
\begin{eqnarray}
P=\begin{pmatrix}
1&0&0\\
0&-1&0\\
0&0&-1
\end{pmatrix},
\end{eqnarray}
only the component $\chi^\nu_1$ has the zero-mode.
Furthermore, when
\begin{eqnarray}
R=\begin{pmatrix}
0&\rho^\ell &0\\
0&0&\rho^m \\
\rho^n &0&0
\end{pmatrix},
\end{eqnarray}
where $\ell+m+n=0$ mod $N$, 
only the direction $(\chi_1^\ell,\chi_2^\ell,\chi_3^\ell)
=\chi_1^\ell (1,\rho^i,\rho^j)$, 
where $i,j=0,1,\cdots,N-1$, has the zero-mode.
If these zero modes develop their VEVs in four-dimensional 
effective field theory, the vacuum alignment can be 
realized.

%It is convenient to write this matrix in the diagonal form 
%$(\chi^\ell=U_R \bar\chi^\ell)$
%\begin{eqnarray}
%U_R^\dagger Z_3U_R
%=\begin{pmatrix}
%1&0&0\\
%0&\omega^2&0\\
%0&0&\omega
%\end{pmatrix},
%\quad
%U_R
%=\frac{1}{\sqrt3}
%\begin{pmatrix}
%\rho^{-n}&0&0\\
%0&\rho^{-m}&0\\
%0&0&1
%\end{pmatrix}
%\begin{pmatrix}
%1&\omega^2&\omega\\
%1&\omega&\omega^2\\
%1&1&1
%\end{pmatrix}.
%\end{eqnarray} 
%Then only ${\bar \chi_1}^\nu$ has a zero mode with a flat wavefunction. 
%The vacuum expectation value of original field becomes 
%$\langle(\chi_1^\ell,\chi_2^\ell,\chi_3^\ell)\rangle
%=\langle\chi_1^\ell\rangle(1,\rho^i,\rho^j)$, 
%where $i,j=0,1,\cdots,N-1$. 

\section{Higher order corrections}

Let us consider higher dimensional terms for 
charged leptons. We do not consider terms with 
$(\chi_0^\nu)^2$ or $(\chi_1^\nu,\chi_2^\nu,\chi_3^\nu)^2$ 
because it just changes the mass eigenvalues. 
Next-to-leading terms which modify eigenvectors are 
\begin{eqnarray}
\begin{split}
\Delta w_e
=e^c(\ell_e,\ell_\mu,\ell_\tau)H_d
(\Delta y^e_1 \chi_0^\nu(\chi_1^\nu,\chi_2^\nu,\chi_3^\nu)
+\Delta y^e_2 (\chi_1^\nu,\chi_2^\nu,\chi_3^\nu)^2)
(\chi_1^\ell\chi_2^\ell\chi_3^\ell)^{N/2+a}
/\Lambda^{3N/2+3a+2}
\\
+ \mu^c(\ell_e,\ell_\mu,\ell_\tau)H_d
(\Delta y^\mu_1 \chi_0^\nu(\chi_1^\nu,\chi_2^\nu,\chi_3^\nu)
+\Delta y^\mu_2 (\chi_1^\nu,\chi_2^\nu,\chi_3^\nu)^2)
(\chi_1^\ell\chi_2^\ell\chi_3^\ell)^{N/2+b}
/\Lambda^{3N/2+3b+2}
\\
+ \tau^c(\ell_e,\ell_\mu,\ell_\tau)H_d
(\Delta y^\tau_1 \chi_0^\nu(\chi_1^\nu,\chi_2^\nu,\chi_3^\nu)
+\Delta y^\tau_2 (\chi_1^\nu,\chi_2^\nu,\chi_3^\nu)^2)
(\chi_1^\ell\chi_2^\ell\chi_3^\ell)^{N/2}
/\Lambda^{3N/2+2}.
\end{split}
\end{eqnarray}
Compared to the leading order, 
corrections are of ${\cal O}(\alpha_\ell^N\alpha_\nu^2)$. 

Similarly, correction terms for neutrinos are
\begin{eqnarray}
\begin{split}
\Delta w_\nu
=&(\ell_e,\ell_\mu,\ell_\tau)(\ell_e,\ell_\mu,\ell_\tau)
(\Delta y_{1}^\nu (\chi_1^\nu,\chi_2^\nu,\chi_3^\nu)
+\Delta y_{2}^\nu \chi_0^\nu)
(\chi_2^\ell\chi_3^\ell,\chi_1^\ell\chi_3^\ell,\chi_1^\ell\chi_2^\ell)^{N/2} 
H_uH_u/\Lambda^{N+2}.
\end{split}
\end{eqnarray}
Corrections are of ${\cal O}((\alpha_\ell)^{N})$ 
to the leading terms. This can be estimated as 
$(m_\tau/y^\tau v_d)^2$. If $\tan\beta$ is not 
large, the corrections can be suppressed.

\section{Conclusion}

We have studied the models with the lepton flavor symmetry
$(Z_N \times Z_N \times Z_N)\rtimes Z_3$ 
in order to extend the tri-bimaximal mixing.
The tri-bimaximal mixing can be derived by residual 
$Z_3$ and $Z_2$ symmetries for charged lepton and 
neutrino mass matrices, respectively. 
We extend the symmetry of charged leptons 
to $Z_3$ and $Z_N$ symmetries, then non-vanishing $\theta_{13}$ 
can be obtained. 
For example, when we choose $N=8$, 
our model predicts $\sin\theta_{13}\approx 0.107$, 
$\sin\theta_{12}\approx 0.337$, 
$\sin\theta_{23}\approx 0.651$, and $\delta_{\rm CP}=0$. 
For larger values of $N$, we predict other sets of mixing angles, 
but in any case, our model predicts $\delta_{\rm CP}=0$.

We have also predicted the mass spectrum of neutrinos. 
Three neutrino masses are expressed by two complex 
parameters, so that they are correlated. In our model, 
only normal mass hierarchy is allowed. 
To compare with future experiments, the relation of 
effective mass of double beta decay and minimum neutrino mass 
are calculated.

%%%%% acknowledgement %%%%%
\vspace{1cm}
\noindent
{\bf Acknowledgement}

H.I. is supported by Grand-in-Aid for Scientific Research,
No.23.696 from the Japan Society of Promotion of Science. 
T. K. is supported in part by the Grant-in-Aid for Scientific 
Research No. 20540266 and the
Grant-in-Aid for the Global COE Program ``The Next 
Generation of Physics, Spun from
Universality and Emergence'' from the Ministry of Education, 
Culture,Sports, Science and Technology of Japan.

\end{document}